\documentstyle[12pt]{article}
\newcommand{\beq}{\begin{eqnarray}}
\newcommand{\eeq}{  \end{eqnarray}}

\newcommand{\dv}[2]{\frac{\textstyle #1}{\textstyle #2}}

\def\prd#1 { Phys.\ Rev.\ D {\bf #1 }}
\def\npb#1 { Nucl.\ Phys.\ B {\bf #1 }}
\def\plb#1 { Phys.\ Lett.\ B {\bf #1 }}
\def\zpc#1 { Z.\@ Phys.\ C {\bf #1 }}
\def\prl#1 { Phys.\ Rev.\ Lett.\ {\bf#1 }}

\newcommand{\ra}{\rightarrow}

\newcommand{\vv}[1]{ {\bf #1}}
\def\lp{\lambda^{\prime} }

\def\ga{ \gamma_5    }
\def\an{ {J,J_3} }

\def\ji{ dx\ d^2\vv{k_\bot} }

\def\ci{ {\cal I} }

\begin{document}
\baselineskip=8mm
\newcount\sectionnumber
\sectionnumber=0

\begin{flushright}{ UTPT--94--02 }
\end{flushright}

\vspace{8mm}
\begin{center}
{\bf {\huge Semileptonic and Exclusive Rare B Decays} }\\
\vspace{6mm}
Patrick J. O'Donnell and Q. P. Xu\\
Physics Department,\\
University of Toronto,\\
Toronto, Ontario M5S 1A7, Canada.\\
\end{center}

\bigskip

The exclusive  rare decay $B \ra K^\ast \gamma$
takes place in a region of maximum  recoil,
$q^{2}=0$,  posing a problem  for  nonrelativistic  quark  models
which are  usually  thought to be most  reliable at zero  recoil.
The   Bauer--Stech--Wirbel   (BSW)   model,   formulated   in   the
infinite--momentum--frame  (IMF)  formalism, is designed to work at
$q^2=0$.  We show in this  model  that  the  ratio  relating  the
decay $B \ra K^\ast \gamma$ and the $q^2$--spectrum
of the  semileptonic  decay  $B\ra  \rho e {\bar  \nu}$,  becomes
independent  of the wave  function in the SU(3)  flavor  symmetry
limit.  We show that this  feature  is also true in  relativistic
quark models  formulated in the IMF or  light--cone  formalism, if
the  $b$ quark is  infinitely  heavy.  In fact, these  relativistic
models, which have a different spin  structure from the BSW case,
reduce  to the BSW  model in the  heavy $b$--quark  limit.  A direct
measurement of the  $q^2$--spectrum of the semileptonic  decay can
therefore  provide  accurate  information  for the exclusive rare
decay.

\newpage

\section{Introduction}

The first experimental observation of the exclusive decay $B
\ra K^\ast$ has been reported from the CLEO collaboration
\cite{CLEO} which gives a branching ratio of
$(4.5\pm 1.5\pm 0.9)\times 10^{-5}$. Theoretically, this rare
decay is not well understood, as there is still an uncertainty
of a factor of about 10 in the branching ratio depending on the
way the large recoil of the $K^\ast$ is handled \cite{OT0} in
the form factors. Burdman and Donoghue \cite{BD} pointed out
that the heavy--quark symmetry together
with the SU(3) flavor symmetry
could relate the rare decay $B\rightarrow K^{\ast}\gamma$ to a
measurement of the semileptonic decay $B\rightarrow\rho
e\bar{\nu}$, independent of the form factors. However, the
relation is only valid at a single point in the Dalitz plot, a
point where the semileptonic decay vanishes, so that there
would still be a large uncertainty in such a measurement. In a
recent paper \cite{OT} O'Donnell and Tung studied the
possibility of relating the decay $B\rightarrow K^{\ast}\gamma$
to the $q^{2}$--spectrum of the semileptonic decay
$B\rightarrow\rho e\bar{\nu}$. They showed that the ratio $\ci$
(to be defined below), which relates the two decays at
$q^{2}=0$, is quite insensitive to different models and the
wave functions. Since the $q^2$--spectrum of
$B\rightarrow\rho e\bar{\nu}$ does not vanish at $q^{2}=0$, a
direct measurement of the spectrum at this point can therefore
provide quite accurate information for $B\rightarrow
K^{\ast}\gamma$. An application of this was made in Ref.\ \cite{Sa}.

In  this   letter  we  first   study  the  ratio   $\ci$  in
the BSW model  \cite{BSW}.  We find that
in the SU(3) flavor  symmetry  limit, this ratio becomes
independent of the  wave  function   used  in  this  model.
The  BSW  model  is formulated in the
IMF  formalism and, since $q^{2}=0$ is the maximum recoil
region, this seems to be an appropriate frame.  The model is
not a completely  relativistic  one, however, in the sense
that the spin and  orbital  parts of the wave  function are
factorized.  There are relativistic quark models formulated in
the IMF or light--cone formalism  \cite{ru1,ru2}  for which the
spin and  orbital  parts of the wave  function are not
factorized.  We show that in such models, the ratio $\ci$ is
also  independent  of the wave  function, if the $b$ quark is
infinitely  heavy.  In fact the relativistic  quark models
become  equivalent to the BSW model in the  heavy  $b$--quark
limit.  This result does not depend on assuming a heavy $s$
quark.

\section{ $B \ra K^\ast\gamma$ versus
$B \ra \rho e \bar \nu$}

We define the form factors in
$B \ra K^\ast\gamma$ and
$B \ra \rho e \bar \nu$ by
\begin{eqnarray}
\langle V(p_V,\epsilon)|\bar{Q}
i\sigma_{\mu\nu}q^{\nu}b_{R}|B(p_{B})\rangle\hspace{-0.6cm}
&&=f_{1}(q^{2})i\varepsilon_{\mu\nu\lambda\sigma}
\epsilon^{\ast\nu}p^{\lambda}_{B} p_V^{\sigma}\nonumber\\
&&+ \left[(m^{2}_{B}-m^{2}_{V})\epsilon^{\ast}_{\mu}-
(\epsilon^{\ast}\cdot q)(p_{B}+p_V)_{\mu}\right]f_{2}(q^{2})\nonumber\\
&&+(\epsilon^{\ast}\cdot q)\!
\left[(p_{B}-p_V)_{\mu}-\frac{q^{2}\ (p_{B}+p_V)_{\mu}}{(m^{2}_{B}
-m^{2}_{V})}\right]f_{3}(q^{2}) ,\nonumber\\
\label{eb1}
\end{eqnarray}
and
\begin{eqnarray}
&&\hspace{-1.2cm}\langle
V(p_V,\epsilon) \mid {\bar Q} \gamma_\mu \gamma_5 b \mid B(p_B)
\rangle
=\dv{2 V(q^2)}{m_B+m_V} i \epsilon_{\mu\nu\alpha\beta}
\epsilon^{\ast\nu} p_B^\alpha p_V^\beta +2m_V
\frac{(\epsilon^\ast \cdot p_B) }{q^2} q^\mu A_0(q^2)\nonumber\\
&&\hspace{-1.2cm}+\!\left[ (m_B+m_V) \epsilon^{\ast\mu} A_1(q^2)\!-
\frac{(\epsilon^\ast \cdot p_B) }{m_B+m_V}(p_B+p_V)^\mu A_2(q^2)\!
-2m_V \frac{(\epsilon^\ast \cdot p_B) }{q^2} q^\mu A_3(q^2)\right]\!\!.
\label{eb2}
\end{eqnarray}

The branching ratio for the exclusive $B\rightarrow K^{\ast}\gamma$
to the inclusive $b\rightarrow s\gamma$ processes
can be written in terms of $f_{1}(0)$ and $f_{2}(0)$ at $q^{2}=0$,
as \cite{Alt,Desh}
\begin{eqnarray}
R(B\rightarrow K^{\ast}\gamma ) &=&
\frac{\Gamma (B\rightarrow K^{\ast}\gamma)}
{\Gamma (b\rightarrow s\gamma)}
\cong \frac{m_{b}^{3}(m_{B}^{2}-m_{K^{\ast}}^{2})^{3}}
{m_{B}^{3}(m_{b}^{2}-m_{s}^{2})^{3}}\frac{1}{2}\left[\,|f_{1}(0)|^{2}
+4|f_{2}(0)|^{2}\,\right].\label{ratio}
\end{eqnarray}
In most models $f_{2}(0)=\dv{1}{2}f_{1}(0)$\footnote{
In the heavy $b$ quark limit \cite{OT} this relation is exact. }.
Although there is now only one form factor to calculate
in Eq.\ (\ref{ratio}), this is still a controversial model-dependent
calculation \cite{OT0}.

Burdman and Donoghue \cite{BD}
give a method of relating $B\rightarrow K^{\ast}\gamma$ to
the semileptonic process
$B\rightarrow \rho e \bar{\nu}$ using the static $b$-quark
limit and SU(3) flavor symmetry, independent of the model
dependence of the form
factors.
Their main result is
\begin{eqnarray}
\Gamma(B\rightarrow K^{\ast}\gamma)\left( \lim_{q^2\rightarrow 0, curve}
\frac{1}{q^2} \frac{d\Gamma (B \rightarrow \rho e \bar{\nu})}{dE_{\rho}dE_e}
\right) ^{-1}
&=&\frac{4 \pi^2}{G^{2}_{F}}\frac{|\eta|^2}{|V_{ub}|^{2}}
\frac{(m_{B}^{2}-m^{2}_{K^{\ast}})^3}{m_{B}^{4}}\ .
\label{BDeq}
\end{eqnarray}
Here $\eta$ represents the QCD corrections to the decay $b
\rightarrow s \gamma$ and the word ``curve'' denotes the
region in the Dalitz plot where $q^2 = 4 E_e (m_B - E_{\rho} - E_e)$.
Their method replaces the uncertainty in the
calculation at large recoil ($q^{2}=0$) of the
$B\rightarrow K^{\ast}$ form factors by making a direct measurement of
$B\rightarrow \rho e\bar{\nu}$. The problem
is that the semileptonic decay vanishes at the $q^2=0$ point on
the ``curve," which is why this kinematic factor is divided out
in Eq.\ (\ref{BDeq}). This means that experimentally there
should be no events at that point and very few in the
neighborhood, making it a very difficult measurement.

To avoid this O'Donnell and Tung \cite{OT} studied
instead the $q^2$--spectrum of the semileptonic
decay $B\rightarrow \rho e \bar{\nu}$.
In this case
\begin{eqnarray}
\hspace{-2cm}
R(B\rightarrow K^{\ast}\gamma )\hspace{-0.5cm}&&
\left( \left. \frac{d\Gamma (B \rightarrow \rho e \bar{\nu})}{dq^2}
\right|_{q^2=0}\right) ^{-1}\nonumber\\
&&=\frac{192\pi^{3}}{G^{2}_{F}} \frac{1}{|V_{ub}|^{2}}
\frac{(m^{2}_{B}-m^{2}_{K^{\ast}})^{5}}{(m^{2}_{B}-m^{2}_{\rho})^{5}}
\frac{(m_{B}-m_{\rho})^{2}}{(m_{B}-m_{K^{\ast}})^{2}}
\frac{m^{3}_{b}}{(m^{2}_{b}-m^{2}_{s})^{3}} \,|{\cal I}|^{2}\ ,
\label{e99}
\end{eqnarray}
where $\ci$ is defined as
\beq
\ci=\dv{ (m_B+m_\rho) }{ (m_B+m_{K^\ast})}
\dv{ f_1^{B\ra K^\ast}(0)}{ A_3^{B\ra \rho}(0) } \ .
\label{e10}
\eeq
The advantage here is
that the $q^2$--spectrum does
not vanish at $q^{2}=0$.
The disadvantage is that in taking the ratio we do not have
in general the simple
cancellation of form factors. However, as we will see, in a number
of models the ratio
$\ci$ is still relatively free of uncertainties.

\section{The BSW model}

The BSW model,  formulated in the IMF  formalism,  is designed to
work at  $q^2=0$  where the ratio  $\ci$ of Eq.\  (\ref{e10})  is
defined.  In this model, the form factors  $f_1(0)$  and $A_3(0)$
are given by
\beq
A_3(0)=f_1(0)= \int\ji\phi^{\ast}_{V}(x,\vv{k_\bot})
\phi_{B}(x,\vv{k_\bot})\,\,\,  ,
\label{g1}
\eeq
where  $\phi(x,\vv{k_\bot})$  is the  wave  function.  Thus,  the
ratio  $\ci$ in Eq.\  (\ref{e10})  becomes  1 if we use the SU(3)
flavor  symmetry for $\rho$ and $K^\ast$.  This is independent of
the wave functions used.

In the limit when both $m_b \ra \infty$ and $m_Q \ra \infty$,
the heavy quark symmetry \cite{IW1} gives $A_3(0)=f_1(0)$ for $
B(\bar q b)\ra  V(\bar q Q)$  transitions; this  equality  is
not  expected  to be  true  for arbitrary $m_b$ and $m_Q$.  In
the BSW model this equality between  $f_1(0)$ and
$A_3(0)$ for arbitrary  $m_b$ and $m_Q$ comes from the
factorization of the spin and orbital parts of the wave
function.  The total wave  function
$\Psi^\an(\vv{p_1},\vv{p_2}, \lambda_1,\lambda_2)$ has the
simple form
\beq
\Psi^\an(\vv{p_1},\vv{p_2}, \lambda_1,\lambda_2)
=\chi^\dagger_{\lambda_1}
\ S^\an\ \chi_{\lambda_2}\ \phi(x,\vv{k_\bot})\ ,
\label{e25}
\eeq
where $\lambda_{1,2}$ are the spin indices of the quarks and
$\chi$ is the Pauli spinor.
The relation between the coordinates $\vv{p_{1,2}}$ and
$(x,\vv{k_\bot})$ is given in Eq.\ (\ref{e32}) below.
For the pseudoscalar and vector mesons,
the spin wave functions are given by the SU(2) relations
\beq
S^{0,0}=\dv{i \sigma_2}{\sqrt2} \ \ , \ \
S^{1,\pm 1}=\dv{1\pm\sigma_3}{2}\ \ , \ \
S^{1,0}=\dv{\sigma_1}{\sqrt2} \ .
\label{e27}
\eeq
Hence, there is a spin symmetry in the model which gives
simple  relations  among the form factors.  For example, the form
factors  $A_1(0)$  and  $V(0)$  are also  related \cite{BSW}.

\section{Relativistic quark models with IMF formalism}

Due to its  treatment  of the quark spins, the BSW model is not
a completely  relativistic  quark  model.  A  method  to
obtain  a relativistic  quark model using the IMF or
light--cone  formalism was developed quite a long time ago
\cite{ru1,ru2} and there have been many applications
\cite{ru1,ru2,chung,jaus,schlumpf}.  In \cite{ru2} and
\cite{chung}, the model was applied to calculate
electromagnetic  form  factors of pions and nucleons.  In
\cite{jaus} and \cite{schlumpf}  similar models were used to
study weak  decays of mesons and  baryons.

We describe a ground-state $Q{\bar q}$
meson V in the infinite momentum frame by
\beq
| V (P, J_3, J) \rangle=\!\!\!\!\!&&\!\!\!\!\!\int d^3\vv{p_1} d^3\vv{p_2}
\ \delta(\vv{P}-\vv{p_1}-\vv{p_2}) \nonumber\\
\!\!\!\!&&\!\!\sum_{\lambda_1,\lambda_2}
\Psi^\an(\vv{P},\vv{p_1},\vv{p_2},\lambda_1,\lambda_2)
|Q(\lambda_1,\vv{p_1} )\ {\bar q}(\lambda_2,\vv{p_2} )\rangle\ ,
\label{e31}
\eeq
where $\vv{P}=P\vv{e}_z, P\ra \infty$ and
the quark coordinates are
\beq
&&p_{1z}=x_1 P \ , p_{2z}=x_2 P \ ,
\ x_1+x_2=1 \ , 0\le x_{1,2}\le 1 \ ,\nonumber\\
&&\vv{p_{1\bot}}= \vv{k_\bot} \ , \
\vv{p_{2\bot}}=-\vv{k_\bot} \ .
\label{e32}
\eeq
Rotational invariance of the wave function for states with spin
$J$ and zero orbital angular momentum requires the wave function
to have the form \cite{chung,jaus} (with $x=x_1$)
\beq
\Psi^{J,J_3}(\vv{P},\vv{p_1},\vv{p_2},\lambda_1,\lambda_2)
=R^\an(\vv{k_\bot},\lambda_1,\lambda_2)
\phi(x, \vv{k_\bot}),
\label{e36}
\eeq
where $\phi(x, \vv{k_\bot})$ is even in $\vv{k_\bot}$ and
\beq
&&R^{J,J_3}(\vv{k_\bot},\lambda_1,\lambda_2)
\!\!=\nonumber\\
&&\sum_{\lambda,\lp}
\langle \lambda_1 | R^\dagger_M(\vv{ k_\bot},m_Q) |\lambda \rangle
\langle \lambda_2 | R^\dagger_M(\vv{-k_\bot},m_{\bar q}) |\lp   \rangle \
C^{J,J_3}(\dv{1}{2},\lambda;\dv{1}{2},\lambda^\prime)\ .
\label{e37}
\eeq
In Eq.\ (\ref{e37}),
$C^{J,J_3}(\dv{1}{2},\lambda;\dv{1}{2},\lambda^\prime)$ is
the Clebsh-Gordan coefficient and the rotation
$R_M(\vv{k_\bot}, m_i)$ ($i=Q, \bar q=1, 2$)
on the quark spins is the Melosh rotation \cite{melosh}:
\beq
R_M(\vv{k_\bot},m_i)=
\dv{m_i+x_i M_0-i
{\bf \sigma}\vv{\cdot}(\vv{n\times k_\bot})}
{\sqrt{ (m_i+x_i M_0)^2+\vv{k}^2_{\bot} } } \ ,
\label{e33}
\eeq
where $\vv{n}=(0,0,1)$ and
\beq
M_0^2=\dv{m_1^2+\vv{k}^2_{\bot}}{x_1}+\dv{m_2^2+\vv{k}^2_{\bot}}{x_2} \ .
\label{e34}
\eeq
It's easy to see that if there is no transverse momentum, the rotation
becomes an unity matrix:
\beq
R_M(\vv{0},m_i)=1 \ .
\label{e35}
\eeq
The spin wave function
$R^{J,J_3}(\vv{k_\bot},\lambda_1,\lambda_2)$ in Eq.\ (\ref{e37}) can also
be written as
\beq
R^{J,J_3}(\vv{k_\bot},\lambda_1,\lambda_2)\!\!\!&&=
\chi^\dagger_{\lambda_1} R^\dagger_M(\vv{ k_\bot},m_Q)
\ S^\an\ R^{\dagger T}_M(\vv{-k_\bot},m_{\bar q}) \chi_{\lambda_2} \nonumber\\
\!\!\!&&=\chi^\dagger_{\lambda_1}\ U^\an_V\ \chi_{\lambda_2} \ ,
\label{e35a}
\eeq
where $S^\an$ is defined by
\beq
S^{J,J_3}=\sum_{\lambda,\lambda^\prime}
|\lambda\rangle \langle \lambda^\prime | \
C^{J,J_3}(\dv{1}{2},\lambda;\dv{1}{2},\lambda^\prime) \ .
\label{e35b}
\eeq
For the pseudoscalar and vector mesons, $S^{J,J_3}$ is given
in Eq.\ (\ref{e27}) and the $U^{J,J_3}$ read
\beq
&&\hspace{-1.4cm}U^{0,0}=\dv{
\sqrt2 (x_1\ m_2+x_2\ m_1)\ S^{0,0}-\vv{k}_-\ S^{1,1}
-\vv{k}_+\ S^{1,-1} }{\sqrt2\ d_0}\ , \nonumber\\
&&\hspace{-1.4cm}U^{1,0}=\dv{
\sqrt2 (\alpha_1\cdot\alpha_2+\vv{k}^2_\bot)\ S^{1,0}
+(m_1-m_2+(x_1-x_2) M_0)
(\vv{k}_+\ S^{1,-1}-\vv{k}_-\ S^{1,1})
}{\sqrt2\ d_1 d_2}\ ,\nonumber\\
&&\hspace{-1.4cm}U^{1,1}=\dv{
 \sqrt2 \alpha_1\cdot \alpha_2\ S^{1,1}
-\sqrt2\ \vv{k}^2_+\ S^{1,-1}
+(\alpha_1-\alpha_2) \vv{k}_+\ S^{1,0}
+(\alpha_1+\alpha_2) \vv{k}_+\ S^{0,0}
}{\sqrt2\ d_1 d_2}\ , \nonumber\\
&&\hspace{-1.4cm}U^{1,\!-1}\!=\!\dv{
-\sqrt2\ \vv{k}^2_-\ S^{1,1}
+\sqrt2 \alpha_1\cdot \alpha_2\ S^{1,-1}
-(\alpha_1-\alpha_2) \vv{k}_-\ S^{1,0}
+(\alpha_1+\alpha_2) \vv{k}_-\ S^{0,0}
}{\sqrt2\ d_1 d_2} ,
\label{e35c}
\eeq
where
\beq
\alpha_1=m_1+x_1 M_0\ , \;\alpha_2=m_2+x_2 M_0\ ,
\vv{k}_\pm=\vv{k}_x\pm i \vv{k}_y\ ,
\label{new1}
\eeq
and
\beq
&&d_0=\sqrt{\vv{k}^2_\bot+(x_1 m_2+x_2 m_1)^2}
\ ,\ \ d_1=\sqrt{ (m_1+x_1 M_0)^2+\vv{k}^2_\bot }\ ,\nonumber\\
&&d_2=\sqrt{ (m_2+x_2 M_0)^2+\vv{k}^2_\bot }\ .
\label{new2}
\eeq
The normalization condition for
the wave function is
\beq
\hspace{1cm}
1=\int dx\ d^2\vv{k_\bot} \sum_{\lambda_1,\lambda_2}
|\Psi^{J,J_3}(\vv{p_1},\vv{p_2},
\lambda_1,\lambda_2) |^2
=\int dx\ d^2\vv{k_\bot} |\phi(x, \vv{k_\bot})|^2\ ,
\label{e35d}
\eeq
where
\beq
R^\dagger_M(\vv{k_\bot},m_i) R_M(\vv{k_\bot},m_i)=1 \ .
\eeq

The matrix element of a $B$ meson decaying to a vector meson $V$
is
\beq
\!\!\!\!\langle V(p_V,J_3)|\bar{Q}
\Gamma b|B(p_B) \rangle \!\!\!\!
&&\!\!\!\!=P \int dx\ d^2\vv{k_\bot}
\sum_{\lambda_1,\lambda_2}
\dv{ \Psi^{\ast 1,J_3}_V\ \bar u_Q \Gamma u_b\ \Psi_B }
{\sqrt{ 4 e_b e_Q } }
\nonumber\\
\!\!\!\!&&\!\!\!\!= P \int dx\ d^2\vv{k_\bot}
\dv{ \phi^\ast_V \phi_B }{\sqrt{ 4 e_b e_Q } }
Tr\left[ U^{\dagger 1,J_3}_V\ U_\Gamma\ U^{0,0}_B \right]\ ,
\label{e38}
\eeq
where $U_\Gamma$ is defined by
\beq
{\bar u}^i_Q\ \Gamma\ u^j_b=\chi^\dagger_i\ U_\Gamma\ \chi_j \ ,
\label{e39}
\eeq
and $e_b$ and $e_Q$ are energies of the quarks.
In Eq.\ (\ref{e38}), we choose $\vv{p_B}=P \vv{e_z},\ P\ra \infty $,
and the momentum transfer $q=p_B-p_V$ is given by
\beq
\vv{q_\bot}=0, \ q_0=-q_z=\dv{(m_B^2-m_V^2)}{4 P}\ , \ q^2=0\ .
\label{e39b}
\eeq
In contrast to Eq.\ (\ref{e38}),
the matrix element in the nonrelativistic treatment of
quark spins is given by
\beq
\langle V(p_V,J_3)|\bar{Q}
\Gamma b|B(p_B) \rangle
\!\!= P\int \ji
\dv{ \phi^\ast_V \phi_B }{\sqrt{ 4 e_b e_Q } }
Tr\left[ S^{\dagger 1,J_3} U_\Gamma S^{0,0} \right]   \ .
\label{e310}
\eeq

Using Eq.\ (\ref{e38}), we obtain new expressions for the
form factors $A_3(0)$ and $f_1(0)$
\beq
A_3(0)&=&\int dx\ d^2\vv{k_\bot} \phi^\ast_V \phi_B\ \dv{T_{A_3}}{D}\ ,
\nonumber\\
f_1(0)&=&\int dx\ d^2\vv{k_\bot} \phi^\ast_V \phi_B\ \dv{T_{f_1}}{D}\ ,
\label{a1}
\eeq
where the kinematic terms are
\beq
T_{A_3}&=&\left[ (m_Q+x M^V_0)(m_{\bar q}+(1-x)M^V_0)+\vv{k}^2_{\bot} \right]
(x\ m_{\bar q}+(1-x)m_b) \nonumber\\
&&+(m_Q-m_{\bar q}+(2x-1) M^V_0 ) \vv{k}^2_{\bot}\ ,
\nonumber\\
T_{f_1}&=&(m_Q+x M^V_0)\left[ \vv{k}^2_{\bot}+
(x\ m_{\bar q}+(1-x) m_b)(m_{\bar q}+(1-x) M^V_0)\right]\ ,\\
D&=&\sqrt{\vv{k}^2_{\bot}+(x\ m_{\bar q}+(1-x) m_b)^2}\nonumber \\
&& \times
\sqrt{ (m_Q+x M^V_0)^2+\vv{k}^2_{\bot} }
\sqrt{ (m_{\bar q}+(1-x) M^V_0)^2+\vv{k}^2_{\bot} }\ .
\label{a2}
\eeq
Here, $M^V_0$ corresponds to Eq.\ (\ref{e34}) for the meson V.

One interesting feature of the transformation Eq.\ (\ref{e37})
is that for the heavy pseudoscalar mesons, such as the B meson, one can write
\beq
R^{0,0}(\vv{k},\lambda_1,\lambda_2)=
\dv{ {\bar u}_b(\vv{p_1}, \lambda_1)
\ga v_{\bar q}(\vv{p_2}, \lambda_2)}
{\sqrt2 \sqrt{M_0^2-(m_b-m_{\bar q})^2}}\ ,
\label{e318}
\eeq
and for the heavy vector mesons such as the $B^\ast$ meson
\beq
R^{1,J_3}(\vv{k},\lambda_1,\lambda_2)=
-\dv{\zeta^\mu(J_3) {\bar u}_b(\vv{p_1}, \lambda_1) \gamma_\mu
 v_{\bar q}(\vv{p_2}, \lambda_2)}
{\sqrt{M_0^2-(m_b-m_{\bar q})^2}}\ .
\label{e319}
\eeq

If both the $V$ and B mesons in Eq.  (\ref{e38})  are  heavy, one
can express matrix element in terms of the trace  expression  (as
in the heavy quark effective theory \cite{bj,fggw})
\beq
\langle V(v^{\prime},J_3)|\bar{Q}
\Gamma b|B(v) \rangle
\ \propto
\ Tr\left[ \dv{1+\not\!{v^\prime}}{2} \not\!\epsilon
\ \Gamma \dv{1+\not\!{v}}{2}\ga \right]
\label{e310b}
\eeq

The  meson  wave   functions   $\phi(x,\vv{k_\bot})$   are  model
dependent  and  difficult  to  obtain;  often  simple  forms  are
assumed  for them.  One  possibility  is a Gaussian
type of wave function \cite{ru2,huang}
\beq
\phi(x,\vv{k_\bot})
=N exp\left( -\dv{M^2_0}{2 \beta^2} \right)
=N exp\left( -\dv{1}{2 \beta^2}
\left[ \dv{m_1^2+\vv{k}^2_{\bot}}{x}+
\dv{m_2^2+\vv{k}^2_{\bot}}{1-x} \right] \right)\ .
\label{e311}
\eeq
In \cite{chung,jaus} a slightly
different harmonic-oscillator wave function $\eta(\vv{k})$ was used,
\beq
\eta(\vv{k})=N exp\left(-\dv{\vv{k}^2}{2 \sigma^2} \right) \ ,
\label{e312a}
\eeq
with the normalization,
\beq
1=\int \ji |\phi(x,\vv{k_\bot})|^2=
\int d^3\vv{k} |\eta(\vv{k})|^2.
\label{e312b}
\eeq
Here, $\vv{k}_z$ is defined by $x_1 M_0=E_1+k_z$ with
$E_1=\sqrt{ \vv{k}^2_{\bot}+k^2_z+m_1^2}$ so that
\beq
\phi(x,\vv{k_\bot})=\sqrt{\dv{d\vv{k}_z}{dx} } \eta(\vv{k}).
\label{e312c}
\eeq
A third possibility is the wave function from a relativistic
harmonic oscillator equation \cite{BSW}
\beq
\phi(x,\vv{k_\bot})=
N \sqrt{x(1-x)}
exp\left( -\ \dv{m_B^2}{2w^2}
\left[ x-\dv{1}{2}-\dv{m_b^2-m_q^2}{2m_B^2}\right]^2\right)
\dv{ exp\left(-\ \dv{ \vv{k}^2_{\bot} }{2w^2}\right)}
{\sqrt{\pi w^2} } \ .
\label{e313}
\eeq
The parameters $\beta$, $\sigma$ and $\omega$ in
Eq.\ (\ref{e311}), Eq.\ (\ref{e312a}) and Eq.\ (\ref{e313})
are all of the order of  $\Lambda_{\rm QCD}$.

In the heavy $b$--quark limit, it is well known that the
distribution amplitude
$\int d^2\vv{k}_\bot \phi(x,\vv{k}_\bot)$ of a heavy meson
such as the B meson, has a peak near
$x\simeq x_0=\dv{m_b}{m_B}$.
As the $b$ quark mass  becomes  larger,  the
width of the peak decreases and $x_0$ comes closer to 1.
For the wave function $\phi(x,\vv{k}_\bot)$ itself,
one expects a similar picture: $\phi(x,\vv{k}_\bot)$ vanishes
if $\vv{k}^2_\bot \gg \Lambda^2_{\rm QCD}$
and peaks as $x\ra 1$. All the three wave functions listed before have
this feature. This is easy to understand
since $\langle \vv{k}^2_\bot \rangle\sim\Lambda^2_{\rm QCD}$. Also, the
average velocity of the heavy quark equals that of the heavy
meson so that the average $x\ra 1$.

The  effect of the above  feature on the  matrix  element  of the
current is  that  the  integrand  in Eq.\ (\ref{e38})  vanishes
everywhere  in the heavy $b$--quark limit,  except for $x \ra 1$ and
small  $\vv{k_\bot}$.  When $x \ra  1$,  for  both  $B$  and  $V$
mesons,
\beq
x M_0 \ra \infty\ , \ (1-x) M_0 \ra 0\ ,
\label{e316}
\eeq
and the Melosh rotations become,
\beq
R_M(\vv{k_\bot}, m_{\bar q}) \ra
\dv{m_{\bar q}-i \vv{\sigma\cdot}(\vv{n\times k_\bot})}
{\sqrt{ m_{\bar q}^2+\vv{k}^2_{\bot} } } ,
R_M(\vv{k_\bot}, m_Q) \ra 1 , R_M(\vv{k_\bot}, m_b) \ra 1 .
\label{e317}
\eeq
Consequently,
\beq
U^{1,J_3}_V \ra S^{1,J_3}
\ \dv{m_{\bar q}-i \vv{\sigma\cdot}(\vv{n\times k_\bot})}
{\sqrt{ m_{\bar q}^2+\vv{k}^2_{\bot} } } \ , \
U^{0,0}_B \ra S^{0,0}
\ \dv{m_{\bar q}-i \vv{\sigma\cdot}(\vv{n\times k_\bot})}
{\sqrt{ m_{\bar q}^2+\vv{k}^2_{\bot} } },
\label{318}
\eeq
and Eq.\ (\ref{e38}) becomes Eq.\ (\ref{e310})
\beq
\!\!\!\!\langle V(p_V,J_3)|\bar{Q}
\Gamma b|B(p_B) \rangle \!\!\!\!
&&\!\!\!\!= P \int dx\ d^2\vv{k_\bot}
\dv{ \phi^\ast_V \phi_B }{\sqrt{ 4 e_b e_Q } }
Tr\left[ U^{\dagger 1,J_3}_V\ U_\Gamma\ U^{0,0}_B \right]
\nonumber\\
&&\!\!\!\!\ra P\int \ji
\dv{ \phi^\ast_V \phi_B }{\sqrt{ 4 e_b e_Q } }
Tr\left[ S^{\dagger 1,J_3} U_\Gamma S^{0,0} \right]   \ .
\label{e318b}
\eeq
We conclude that
in the heavy $b$--quark limit the type of relativistic quark models
formulated above reduce to the BSW model.
Thus, in the heavy $b$--quark limit and in SU(3) flavor symmetry,
the ratio $\ci=1$ of Eq.\ (\ref{e10}) remains true
for the type of relativistic quark models \cite{ru1,ru2}
presented here.

Numerically, the mass of $B$ meson is heavy enough to make the
relativistic model very close to the BSW--type model.
If the SU(3)--flavor symmetry is preserved in the wave functions
(with possible breaking coming from the masses) then
$\ci\simeq 1.1$ for all three models.

\section{ Conclusion }

In this letter, we have studied the ratio $\ci$ of Eq.\
(\ref{e10}) suggested in \cite{OT} to supply a better way to
measure the exclusive rare decay $B \ra K^\ast \gamma$.  This
ratio is free of uncertainties in the wave function in the BSW
model in the SU(3) flavor symmetry.  However, the spin treatment
in this model is not relativistic.  We have investigated the
relativistic models \cite{ru1,ru2,chung,jaus}
and shown that these models even with very different
wave functions reduce to the BSW model in the heavy $b$--quark limit.
Thus in such models, the ratio $\ci$ is also 1, in the limits of heavy
$b$--quark and SU(3)--flavor symmetries, independent of uncertainties
in the wave function.  Numerically, using the physical masses and
different wave functions, the ratio $\ci$ still stays close to 1.
Therefore, a direct measurement of the $q^2$--spectrum of the
semileptonic decay can supply accurate information for $B\ra
K^\ast \gamma$.

\newpage

\vspace{.3in}
\centerline{ {\bf Acknowledgment}}
We thank Humphrey K. K. Tung for many useful discussions.
This work was in part supported by the Natural Sciences and Engineering
Council of Canada.

\end{document}